Review paper:

# Functional Magnetic Resonance Imaging Changes and Increased Muscle Pressure in Fibromyalgia: Insights from Prominent Theories of Pain and Muscle Imaging

**Authors:**

Seth Adler [1], Farzan Vahedifard [2], Rachel Akers [3], Christopher Sica [4], Mehmet Kocak [5], Edwin Moore [6], Marc Minkus [7], Gianna Elias [8], Nikhil Aggarwal [9], Sharon Byrd [10], Mehmoodur Rasheed [11], Robert S. Katz *[12]

[1] Department of Diagnostic Radiology and Nuclear Medicine, Rush Medical College. Email: seth_adler@rush.edu
[2] MD, Department of Diagnostic Radiology and Nuclear Medicine, Rush Medical College. Email: farzan_vahedifard@rush.edu
[3] MS, Department of Surgery, Rush Medical College. Email: rachel_a_akers@rush.edu
[4] Ph.D., Technical & Operations Director MRI, Rush Imaging Research Core, email: christopher_sica@rush.edu
[5] MD, Associate Professor, Department of Diagnostic Radiology and Nuclear Medicine, Rush Medical College. Email: mehmet_kocak@rush.edu
[6] Ph.D., Clinical Professor, Technology Entrepreneur Center, Grainger College of Engineering, University of Illinois, Champaign-Urbana, email: egmoore@illinois.edu
[7] MBA, Director of Program Management, Highland Park, Email: marc.minkus@gmail.com
[8] Bioengineering, Grainger College of Engineering, University of Illinois at Urbana Champaign. Email: grelias2@illinois.edu
[9] Oakland University William Beaumont School of Medicine. Email: nikhilaggarwal@oakland.edu
[10] MD, Professor and Chairperson, Department of Diagnostic Radiology and Nuclear Medicine, Rush Medical College. Email: sharon_byrd@rush.edu
[11] MD, Clinical Associate Professor, Carle Illinois College of Medicine University of Illinois, Email: mrasheed@illinois.edu
[12] MD, Professor, Department of Medicine, Rush Medical College, Chicago, USA. Email: rkatzchil@gmail.com

**Abstract**

Fibromyalgia is a complicated and multifaceted disorder marked by widespread chronic pain, fatigue, and muscle tenderness. Current explanations for the pathophysiology of this condition include the Central Sensitization Theory, Cytokine Inflammation Theory, Muscle Hypoxia, Muscle Tender Point Theory, and Small Fiber Neuropathy Theory. The objective of this review article is to examine and explain each of these current theories and to provide a background on our current understanding of fibromyalgia. The medical literature on this disorder, as well as on the roles of functional magnetic resonance imaging (fMRI) and elastography as diagnostic tools, was reviewed from the 1970s to early 2023, primarily using the PubMed database. Five prominent theories of fibromyalgia etiology were examined: 1) Central Sensitization Theory; 2) Cytokine Inflammation Theory; 3) Muscle Hypoxia; 4) Muscle Tender Point Theory; and 5) Small Fiber Neuropathy Theory.

Previous fMRI studies of FMS have revealed two key findings. First, patients with FMS show altered activation patterns in brain regions involved in pain processing. Second, the connectivity between brain structures in individuals diagnosed with FMS and healthy controls is different. Both of these findings will be expanded upon in this paper.

The article also explores the potential for future research in fibromyalgia due to the advancements in fMRI and elastography techniques, such as shear wave ultrasound. Increased understanding of the underlying mechanisms contributing to fibromyalgia symptoms is necessary for improved diagnosis and treatment, and advanced imaging techniques can aid in this process.

I. **Introduction**

Fibromyalgia Syndrome (FMS) is a multifaceted condition characterized by persistent and widespread musculoskeletal pain, localized tender points, fatigue, disrupted sleep, anxiety, depression, cognitive difficulties, and headaches. [1] The diagnosis of FMS is challenging due to the lack of dedicated clinical resources. Furthermore, while the etiology of FMS remains uncertain, increasing evidence suggests that it may involve alterations in the processing of pain signals. [2] In this review, we discuss the background of muscle pathology and pain in the pathophysiology of FMS and the use of functional magnetic resonance imaging (fMRI) and muscle pressure measurements for diagnosis.

The current diagnostic criteria for fibromyalgia, as set by the American College of Rheumatology (ACR), primarily rely on a combination of clinical symptoms and the widespread pain index (WPI) along with the symptom severity (SS) scale. [3] These criteria were updated in 2010 and have since been used as the standard for diagnosing fibromyalgia. [3] To meet the diagnostic criteria for fibromyalgia, a patient typically needs to have:

1. **Widespread Pain Index (WPI):** Widespread pain is assessed by dividing the body into 19 specific regions and asking the patient to report if they have experienced pain in any of these areas during the previous week. The WPI score ranges from 0 to 19 based on the number of areas with pain.

2. **Symptom Severity (SS) Scale:** The SS scale assesses the severity of various symptoms commonly associated with fibromyalgia, including fatigue, waking unrefreshed, and cognitive symptoms. Each symptom is rated on a scale from 0 to 3, with higher scores indicating more severe symptoms. The scores are then summed to calculate the SS scale score, which can range from 0 to 12.

To receive a diagnosis of fibromyalgia according to ACR criteria, a patient typically needs to meet either of the following conditions:

- WPI score of 7 or higher.
- An SS scale score of 5 or higher, or an SS scale score of 3 to 4 along with a WPI score of 9 or higher.

These criteria help clinicians evaluate the extent of pain and associated symptoms to make a fibromyalgia diagnosis, taking into account both the location and severity of the patient's pain and symptoms. However, excluding other potential diagnoses can be difficult in clinical settings due to numerous coexisting conditions (e.g., hypothyroidism, inflammatory rheumatic diseases, systemic lupus erythematosus). [4] These factors often contribute to chronic pain, which is the primary symptom of FMS (Figure 1).

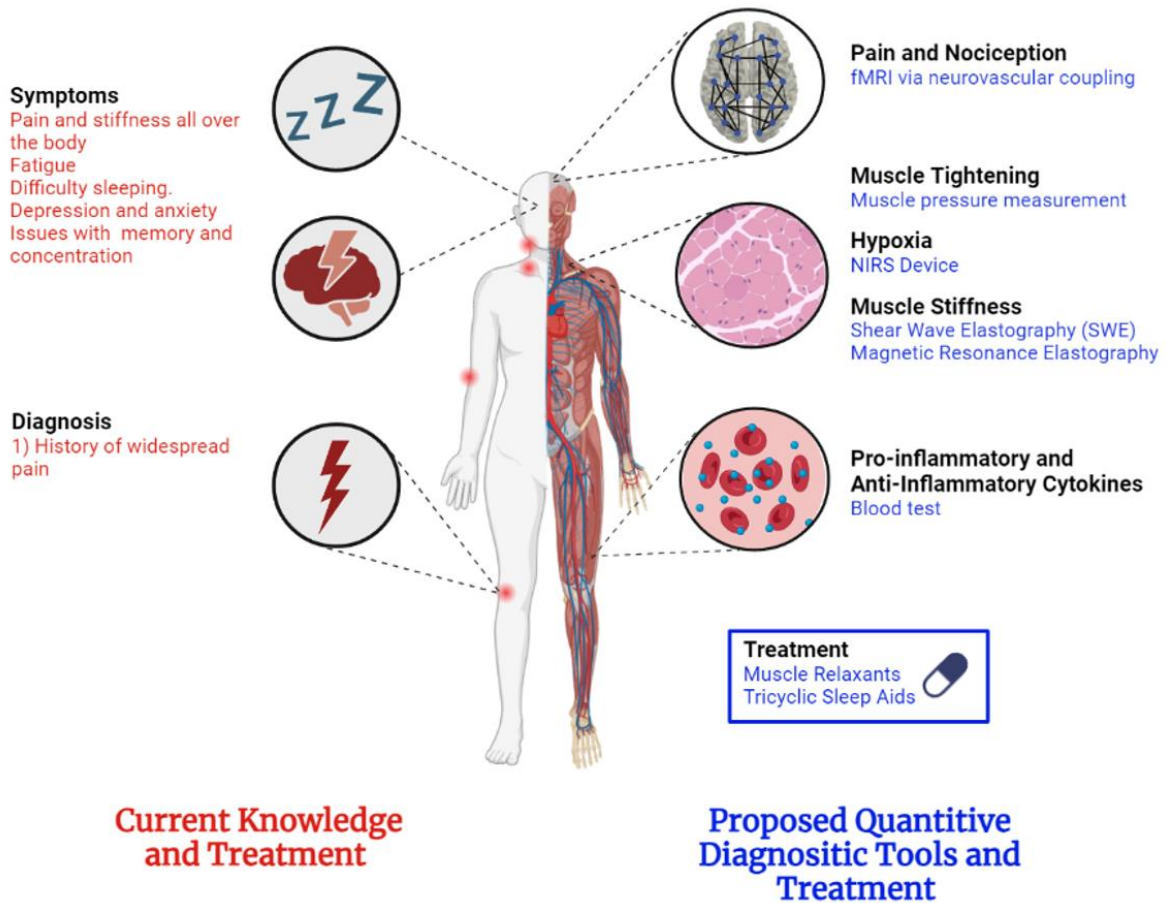

Figure 1  Correlation of increased muscle pressure, pain, fMRI, and lab tests in FMS

II.   **Background**

**Pain**

   To understand FMS, we must first understand the most prevalent and debilitating symptom - pain. The standard definition of pain, established by the International Association for the Study of Pain (IASP), is "an unpleasant sensory and emotional experience associated with, or resembling that associated with, actual or potential tissue damage." [5] The process of nociception consists of noxious signals sent to the peripheral and central nervous systems. This process provides neural feedback that enables the central nervous system to detect and avoid potentially damaging stimuli in active and passive settings. [6]

Pain signals originate from specialized sensory neurons known as nociceptors. When tissue damage or inflammation occurs, these nerve terminals depolarize and transmit "danger" signals to the brain. [7] Based on data by Macpherson et al., [8] it was found that TRPA1 (transient receptor potential ankyrin 1, also known as transient receptor potential cation channel, subfamily A, member 1), an ion channel located on the plasma membrane of many human and animal cells, plays an essential role in propagating the pain response to chemicals that mediate widespread protein modification and tissue damage.TRPA1, also known as the Wasabi Receptor, is activated by reactive chemicals (including cinnamaldehyde, mustard oil, and iodoacetamide) that can form covalent bonds with free cysteine and lysine residues and propagate a pain response.

Assessing musculoskeletal pain due to myalgia is imperative for understanding the pathophysiology of FMS. [9] Myalgia involves mechanisms beyond a single ion channel and is not yet fully understood. Unlike localized skin pain, which is primarily explained by the presence of free nerve endings, myalgia is often diffuse, without localization, and can cause referred pain. [10] The terms used to describe deep tissue pain (e.g., "cramping," "aching") vary among patients and over time, while cutaneous pain is more consistently described as "burning" or "sharp." [11] Treating myalgia is challenging because appropriate pain management is dependent on the underlying cause. First-line therapies for persistent myalgia include nonsteroidal anti-inflammatory drugs (NSAIDs) and physical activity. [12] However, these treatments vary drastically in their mechanism of action and could harm specific patient populations. [13] Therefore, understanding the origin of myalgia in different presentations of illness is crucial to improving the availability and efficacy of targeted pain management strategies, especially in patients with FMS.

**The American College of Rheumatology Classification of Fibromyalgia**

The initial attempt to establish classification criteria for FMS dates back to before 1990 and involved studies conducted by the ACR in 16 centers across the United States and Canada. [14] The results from their initial project are summarized in Tables 1 and 2, which have been reproduced from Wolfe et al.'s 2010 publication in Arthritis Care & Research. [3] The four core areas assessed were pain intensity, physical functioning, emotional functioning, and overall improvement/well-being. [15] Most researchers agreed on the importance of evaluating multiple aspects of FMS, and since the late 1990s, a key focus has been on developing disease-specific measures for each of the four core areas.

Table 1  Fibromyalgia Diagnostic Criteria (Wolfe, 2010)

Criteria

A patient satisfies diagnostic criteria for fibromyalgia if the following 3 conditions are met:

1. Widespread pain index (WPI) ≥ 7 and symptom severity (SS) scale score ≥ 5 or WPI 3-6 and SS scale score ≥ 9.
2. Symptoms have been present at a similar level for at least 3 months.
3. The patient does not have a disorder that would otherwise explain the pain.

Ascertainment

1. WPI: note the number areas in which the patient has had pain over the last week. In how many areas has the patient had pain? Score will be between 0 and 19.

   | Shoulder girdle, left | Hip (buttock, trochanter), left | Jaw, left | Upper back |
   | Shoulder girdle, right | Hip (buttock, trochanter), right | Jaw, right | Lower back |
   | Upper arm, left | Upper leg, left | Chest | Neck |
   | Upper arm, right | Upper leg, right | Abdomen | |
   | Lower arm, left | Lower leg, left | | |
   | Lower arm, right | Lower leg, right | | |

2. SS scale score:
   Fatigue
   Waking unrefreshed
   Cognitive symptoms
   For each of the 3 symptoms above, indicate the level of severity over the past week using the following scale:
   0 = no problem
   1 = slight or mild problems, generally mild or intermittent
   2 = moderate, considerable problems, often present and/or at a moderate level
   3 = severe: pervasive, continuous, life-disturbing problems
   Considering somatic symptoms in general, indicate whether the patient has:
   0 = no symptoms
   1 = few symptoms
   2 = a moderate number of symptoms
   3 = a great deal of symptoms

The SS scale score is the sum of the severity of the 3 symptoms (fatigue, waking unrefreshed, cognitive symptoms) plus the extent (severity) of somatic symptoms in general. The final score is between 0 and 12.

Table 2  Percentage of patients correctly classified according to ACR (American College of Rheumatology) 1990 classification criteria status and diagnostic criteria (Wolfe, 2010)

| Comparison | Phase 1 | | Phase 2 | |
| --- | --- | --- | --- | --- |
|  | Prior FM excluded | All patients | Prior FM excluded | All patients |
| ACR classification criteria positive, FM patients |  | 74.0 |  | 75.6 |
| Classifying patients according to ACR 1990 classification criteria status |  |  |  |  |
| ACR 1990 criteria vs. physician diagnosis |  | 84.1 |  |  |
| ACR 1990 criteria vs. WPI ≥ 7 | 89.0 | 83.6 | 92.3 | 87.9 |
| ACR 1990 criteria vs. WPI ≥ 7 plus muscle pain | 89.9 | 84.6 | 95.2 | 90.5 |
| ACR 1990 criteria vs. WPI ≥ 7 plus muscle tenderness | 92.6 | 87.4 | 95.2 | 90.5 |
| FM diagnostic criteria |  |  |  |  |
| ACR 1990 classification criteria vs. diagnostic criteria ((WPI ≥ 7 AND SS ≥ 5) OR SS ≥ 9) | 88.1 | 82.6 | 95.2 | 90.8 |
| ACR 1990 classification criteria vs. categorical SS scale (SS ≥ 7) | 84.3 | 79.2 | 88.8 | 84.5 |
| Diagnostic criteria ((WPI ≥ 7 AND SS ≥ 5) OR SS ≥ 9) vs. categorical SS scale (SS ≥ 7) |  | 92.3 |  | 89.2 |

In 2010, the ACR introduced new diagnostic criteria for FMS that considered the number of painful body regions and evaluated fatigue, cognitive difficulty, unrefreshed sleep, and the extent of somatic symptoms. [16] The criteria did not use laboratory or radiologic testing, but instead utilized a Symptom Severity (SS) scale ranging from 0 to 12 to quantify the severity of FMS-type symptoms. Later, the SS scale was combined with the Widespread Pain Index (WPI) to create a Fibromyalgianess Scale (FS) ranging from 0 to 31. With a specificity of 96.6% and sensitivity of 91.8%, an FS score ≥ 13 correctly classified 93% of patients identified as having FMS based on the 1990 criteria. [17] The ACR 2010 criteria showed greater sensitivity than the 1990 criteria, enabling the identification of patients with undiagnosed FMS and providing an opportunity for appropriate treatment.

In 2016, an updated revision of the 2010/2011 FMS criteria was developed to create a standardized definition of FMS using measurable aspects. [18] This revision incorporated generalized pain criteria and clinical data. The revised criteria included the following: 1) presence of generalized pain in at least four of five specific regions; 2) symptoms persisting at a similar level for a minimum of three months; 3) a combination of a WPI of 7 or higher and an SS scale of 5 or higher, or a WPI of 4-6 and an SS scale of 9 or higher; and 4) the diagnosis of FMS is considered valid regardless of other concurrent diagnoses. Notably, the presence of other clinically significant illnesses did not exclude the diagnosis of FMS.

As of the publication of this review article, the latest update to the official diagnosis of FMS came in 2019. [19] In collaboration with the World Health Organization (WHO), the IASP Working Group developed a classification system for FMS within the International Classification of Diseases (ICD-11). FMS is now classified as a chronic primary pain, differentiating it from pain arising as a secondary symptom.

### III.    Prominent Theories of Fibromyalgia Pain

Recent research suggests that heightened muscle pressure may contribute to the development of FMS, as elevated muscle tension may exert additional pressure on pain-sensitive muscle receptors and, therefore, trigger a pain response. [20] Moreover, increased muscle pressure induces inflammation, augments the release of pain-inducing substances, and creates a localized hypoxic environment, thereby exacerbating pain and discomfort. [20]

Evolution of the pathogenesis of fibromyalgia pain has focused on five theories: 1) Central Sensitization Theory; 2) Cytokine Inflammation Theory; 3 Muscle Tender Point Theory; 4) Small Fiber Neuropathy Theory; and 5) Muscle Hypoxia.

**Central Sensitization Theory:** According to this theory, pain is caused by the enhanced function of neurons and circuits in nociceptive pathways due to increased membrane excitability and reduced inhibition. [21] Nociceptor inputs can cause a reversible increase in neuron excitability and synaptic efficacy in central nociceptive pathways, known as central sensitization. This results in heightened pain sensitivity, including dynamic tactile allodynia, increased sensitivity to thermal or pressure stimuli, aftersensations, and enhanced temporal summation. In addition to causing pain hypersensitivity, nociceptor stimulation leads to changes in brain activity that are detectable using electrophysiological or imaging techniques.

In simplified terms, the central sensitization theory can be thought of as the brain's pain processing system becoming overly sensitive and "stuck" in a heightened activity state after an injury. The neural pathway responsible for processing pain signals is persistently active and amplified even after the initial injury has healed. This persistent hyperactivity of pain-processing neurons leads to continuous pain sensations beyond those expected in a typical healing process.

Clinical studies have indicated that central sensitization plays a significant role in conditions such as FMS, osteoarthritis, headaches, neuropathic pain, and post-surgical pain. [22] These conditions share similarities in their clinical presentation, response to centrally acting pain relievers, and absence of inflammation or neural damage, suggesting a common underlying mechanism. [22]

One of the initial indications of generalized central sensitization in patients with FMS was observed in a psychophysical study by Gibson et al. [23] This study revealed widespread decreases in thermal and mechanical pain thresholds, along with increased cerebral laser-evoked potentials. Another study, from Sivilotti and Woolf, utilized ketamine to suggest that the FMS pathology may involve an NMDA-dependent component and proposed that tender points could indicate secondary hyperalgesia resulting from central sensitization. [24] Supporting this notion, Arendt-Nielsen et al. conducted a study demonstrating that patients with FMS had lower pressure thresholds and heightened temporal summation in response to muscle stimulation. [25]

Later studies by Staud and Price showed that the maintenance of induced pain in patients with FMS required less frequent stimulation than in normal controls, and concluded that this increased sensitivity contributed to the experience of persistent pain. [26] A later study by the same group showed that the temporal summation of pain and its maintenance were equally elicited from hands and feet, concluding that pain sensitization in these patients was generalized across the entire nervous system. [27]

A sensory testing study, this time by Desmeuleus et al., revealed that patients with FMS, compared to healthy controls, had decreased heat and cold thresholds and reduced pain tolerance. [28] Patients also had a lower nociceptive reflex threshold, which indicated higher central excitability. These findings suggest that the nociceptive reflex threshold could be used as a potential diagnostic measure in identifying patients for whom centrally acting drugs (e.g., pregabalin, duloxetine, and milnacipran) may be therapeutic. Other studies have since confirmed the increased generalized sensitivity (observed as enhanced cortical evoked potentials) in patients with FMS to pressure, thermal, and electrical stimuli. [29]

**Cytokine Inflammation** Cytokines are small, soluble proteins that play an important role in cell signaling and communication in the immune system. [30] These molecules act as molecular messengers that regulate various cellular activities, including immune responses, inflammation, cell growth, and nociception. Increased cytokine activity can lead to increased sensitivity to pain, which is the defining feature of FMS. Depending on their effects, cytokines may be classified as proinflammatory (e.g., IL-1β, IL-6, IL-8, TNFα, and IL-17), or anti-inflammatory (e.g., IL-4, IL-5, IL-10). [31] In patients with FMS, an imbalance between proinflammatory and anti-inflammatory cytokines is thought to contribute to the widespread pain and hypersensitivity that many patients experience. [32] The pathophysiology of FMS involves several proinflammatory cytokines, including IL-1β, IL-6, IL-8, TNFα, and IL-17, and numerous studies have been conducted to explore their potential involvement. The key characteristics of these cytokines and their involvement in FMS are discussed below:

I. <u>Interleukin-I beta (IL-1β)</u>: Macrophages, monocytes, and microglia cells are responsible for producing and releasing IL-1β, with inflammasomes controlling its synthesis. [33] IL-1β, a potent proinflammatory cytokine, enhances pain conduction and transduction through ion channels, and its increased activity is implicated in autoimmune disorders associated with pain, including inflammatory bowel disease, gout, multiple sclerosis, and rheumatoid arthritis. [34] Several studies investigating the association between IL-1β levels and FMS have found increased levels of IL-1β in patients with FMS compared to controls, supporting a potential connection. [32,35,36] Additionally, a study utilizing low-dose naltrexone, which has anti-inflammatory properties at low dosages, showed a reduction in fibromyalgia symptoms and pain, as well as suppression of various cytokines, including IL-1. [37] However, that study had a short duration, a small sample size, and lacked a control group, which limits its general applicability. While some studies have found support for the involvement of IL-1β in FMS, others have not shown any significant relationship. [32]

II. <u>Interleukin-6 (IL-6)</u>: IL-6 is produced and released by macrophages, monocytes, dendritic cells, and mast cells. It plays an important role in both innate and adaptive immune responses by inducing B lymphocytes to produce immunoglobulins. When bound to its receptor (IL-6R), IL-6 can initiate the JAK-STAT3 intracellular signaling pathway, leading to the expression of genes encoding acute-phase proteins. [38] The role of IL-6 in the central nervous system, specifically in the hypothalamus, is thought to contribute to central sensitization and enhance the translation of nociceptive sensory neurons. Studies on pathological pain have shown high levels of IL-6 and its receptors in the dorsal root ganglion and spinal cord. Increased levels of IL-6 and dysregulation in IL-6R signaling have also been linked to disease pathology in those with pathological pain. [39]

III. <u>Interleukin-8 (IL-8)</u>: IL-8 (CXCL8) is produced by immune, endothelial, and airway smooth muscle cells. [40] This cytokine recruits and activates innate immune cells, and mediates pain in chronic inflammatory disorders (e.g., rheumatoid arthritis, inflammatory bowel disease, psoriasis). [41] Studies focusing on IL-8's effects in FMS found increased levels of IL-8 in the plasma and CSF of patients with FMS. [42,43] This finding supports the hypothesis that IL-8, a proinflammatory cytokine that elevates sympathetic system activity (fatigue, hyperalgesia, allodynia), can be a potential cause of FMS symptoms.

IV. <u>Tumor Necrosis Factor-alpha (TNFα)</u>: TNFα is a proinflammatory mediator secreted by macrophages and microglia that modulates pain signaling through binding to TNF receptor 1 (TNFR1). [43] During emotional and physical stress, two common complaints of patients with FMS, substance P and corticotropin-releasing hormone (CRH), two signaling molecules that play a crucial role in the body's stress response, are released and stimulate mast cell secretion of TNFα. In a study by Tsilioni et al., elevated plasma levels of TNFα, substance P, and CRH were found in patients with FMS compared to healthy controls. [44]

V. <u>Interleukin-17 (IL-17)</u>: The IL-17 family consists of six related cytokines (IL-17A-F), with IL-17A being the most extensively studied member. [45] IL-17, like IL-8, plays a crucial role in the inflammatory response by promoting the production of chemokines and granulocyte colony-stimulating factors, which recruit immune cells to infectious sites. Dysregulation of IL-17 has been implicated in the development of autoimmune disorders, including multiple sclerosis, as well as various chronic inflammatory conditions. [46] Based on a study by Pernambuco et al., patients with FMS were found to have elevated plasma levels of IL-17A, as measured via cytometric bead array. [47] The positive correlation between levels of IL-17A and other proinflammatory cytokines in this study further strengthens the hypothesis that inflammatory mechanisms are involved in the symptoms of FMS.

**Muscle Tender Point Theory:** Tender points are areas of heightened sensitivity in muscles, muscle-tendon junctions, bursa, or fat pads that lead to increased pain or soreness when under pressure. [48] Fibromyalgia is typically associated with a widespread distribution of these tender points. In comparison, trigger points are painful areas in muscles that can be felt upon touch and tend to occur in a more limited regional pattern. [48] Their presence is more indicative of myofascial pain syndrome than of FMS. In some cases, both trigger points and tender points may coexist, leading to overlap syndromes. However, the reliability of identifying trigger points among different examiners has been low in most studies, in contrast to the generally consistent identification of tender points. [49]

**Small Fiber Neuropathy Theory:** A more recent theory explaining the cause of FMS symptoms is that increased muscle pressure in areas of pain can compress small sensory nerves, leading to small fiber neuropathy. [50] Small fiber neuropathy is a disorder affecting the peripheral nerves, particularly small somatic and autonomic fibers, that leads to sensory changes (pain, burning, tingling, numbness) and autonomic dysfunction (dry eyes, dry mouth, dizziness, bladder discomfort). Standard tests like electromyogram and nerve conduction studies are unhelpful, and the diagnosis is based on skin biopsy and quantitative sudomotor axon reflex testing. More research is needed in this area to determine the significance of small fiber neuropathy in FMS pain.

Hypoxic Conditions: Muscle hypoxia is also thought to contribute to FMS symptoms because increased muscle tension leads to decreased circulation and deoxygenation. Oxygen starved tissue can lead to myocyte necrosis ("muscle damage") and pain. [51] Hypoxic conditions in muscles can be identified using multiple methods, including changes in metabolic molecules and histological changes in mitochondria. Studies of energy metabolism in muscle cells have shown metabolic changes during ischemia, including decreased adenosine triphosphate (ATP) and creatine phosphate (CP), and increased adenosine monophosphate (AMP) and creatine.

[51] The pathological distribution of mitochondria in localized muscle hypoxia is often described as "moth-eaten" or as "ragged red" fibers, indicating abnormal proliferation. The hypoxic conditions cause damage to oxidative metabolism and result in an increased number of mitochondria as a compensation mechanism. [9] If similar changes are observed in patients with FMS, then hypoxic conditions could be a plausible explanation for FMS symptoms.

Several studies have shown that the muscles of patients with FMS experience hypoxic conditions. In 1973, Fassbender et al. published data that showed swollen capillary endothelial cells in electron microscope images of trapezius muscle cells of patients with FMS. [52] They hypothesized that this finding was due to local hypoxia that caused degenerative changes. In 1986, Lund et al. published a study that used an MDO (Mehrdraht Dortmund Oberfläche) oxygen electrode to evaluate oxygen levels in the trapezius and brachioradialis muscles of patients with FMS. [53] They found abnormal muscle oxygen tension, indicating uneven capillary perfusion, in all patients with FMS compared to healthy controls. However, the cause of the hypoxia was unknown. In 1986, Bengtsson et al. conducted a study that compared the energy metabolism within the painful, tender points of the trapezius muscle in patients with FMS, the non-painful trapezius muscle tissue of patients with FMS, and the trapezius muscle tissue of healthy controls. [51] They found decreased levels of ATP, adenosine diphosphate (ADP), and CP, as well as increased levels of AMP and creatine in the painful muscles of patients with FMS. In the non-painful trapezius muscle of patients with FMS, ATP and PC levels were decreased compared to healthy controls, indicating that metabolic changes are not limited to painful areas. Of note, however, is that the non-tender trapezius muscle tissue from patients with FMS was only obtained from two patients, leading to further opportunities for additional research in this area. Still, their findings support the hypothesis that pain in patients with FMS is of muscular origin and that hypoxia plays an important role in FMS symptoms.

Recent research on pathogenesis of fibromyalgia [16] adds further support to the spotlight on muscle hypoxia. A 2021 study showed that muscle pressure holds a central role in the development of pain. In this study, intramuscular pressure localized to the trapezius was significantly higher in patients with FMS than rheumatic disease controls. [20] This study used a pressure gauge attached to a No. 22 needle and measured using dolorimetry and digital palpation. The mean pressure pain threshold (PPT) score was 14.5 lbs lower in the FMS group than in the control group (8.03 vs 22.54). On digital palpation, patients with FMS had an average score of 2.09 (compared to an average score of 0.47 for the control group). Lower PPT and higher digital palpation scores indicated greater tenderness. Furthermore, trapezius muscle pressure was significantly higher in patients with FMS than in the rheumatic disease controls. The trapezius muscle for patients with FMS was 33.48 mmHg, compared to an average pressure of 12.23 mmHg in controls. Their reasoning about the processes leading to pain is fully in line with the muscle hypoxia hypothesis. Internal fluid pressure is presumed to cause compression of small vessels inside the muscle leading to reduced blood flow and deoxygenation. When oxygen is not sufficiently available to fulfill the need at the tissue level, pain arises.

IV. The Role of fMRI in Studying Fibromyalgia

Functional magnetic resonance imaging (fMRI) is essential in advancing our knowledge of how the brain functions. [54] By analyzing changes in blood oxygen level-dependent (BOLD) signals and comparing them to baseline readings, researchers can determine the areas of the brain that are activated in response to stimuli. Studying the brain at rest using fMRI, a technique known as resting-state fMRI (rs-fMRI), has become increasingly popular. [55] Rs-fMRI explores the synchronized activity between spatially separate brain regions without the need for a specific stimulus.

The discovery of brain resting-state activity significantly changed our understanding of neural dynamics. [55] Resting-state activity refers to synchronized neural oscillations across large-scale networks that occur without external inputs and persist even in the presence of external inputs. This reflects local information integration, as measured by the frequency of content. Resting-state activity is believed to be influenced by an individual's learning and memory history, which contributes to perceptual variability. [56] See Figure 2. In contrast, task-related activity can be seen as a collection of networks synchronizing with each other or components of these networks breaking apart and connecting with others. [57] Resting-state networks have been identified, and specific pain-related networks were observed. These networks demonstrate that the activity of any single voxel is synchronized and embedded within a larger brain network. A few seed voxels can capture a significant portion of the brain activity associated with pain. However, identifying the pain-related network during non-painful resting states raises questions about its sufficiency in pain perception.

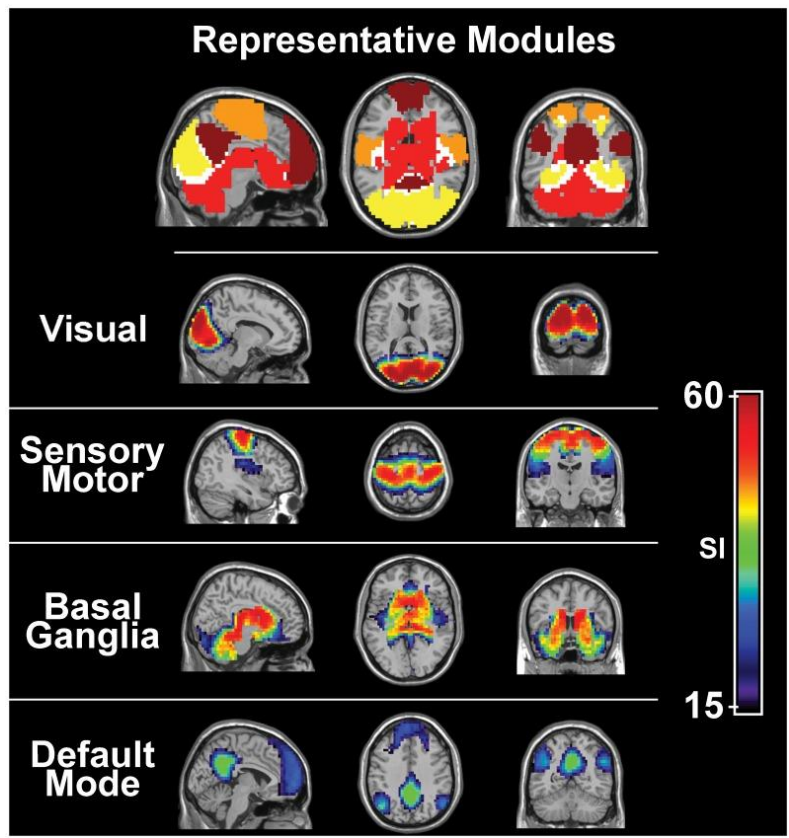

Figure 2  A study identified four consistent functional networks across subjects: visual (yellow), sensory/motor (orange), basal ganglia (red), and default mode (maroon, including posterior cingulate, inferior parietal lobes, and medial frontal gyrus). (Moussa, Malaak N., 2012)

One application of fMRI is in FMS research. To investigate pain processing in patients with FMS, researchers can apply painful stimuli, such as a pinprick or elevated external pressure, to the participant's body while inside the fMRI scanner. By comparing statistical threshold activation maps between patients with FMS and healthy controls, researchers can identify differences in neural responses to pain.

Previous fMRI studies of FMS have revealed several key findings. First, patients with FMS show altered activation patterns in brain regions involved in pain processing. In a study by Aster et al., fMRI was used to highlight the connectivity between brain structures in individuals diagnosed with FMS and healthy controls. [58] They found that those diagnosed with FMS exhibited reduced connectivity between the right mid-frontal gyrus, posterior cerebellum, and right crus cerebellum, and increased connectivity between the inferior frontal gyrus, angular gyrus, and posterior parietal gyrus. These brain regions, some of which are highlighted in Figures 3 and 4, are associated with sensory perception, the cognitive aspects of pain, and pain modulation. In a study by Gracely et al., the group used fMRI to examine cerebral activation patterns during painful pressure application and determine if they differ in patients with FMS compared to controls. [59] Patients with FMS and matched controls underwent fMRI while pressure was applied to the thumb. Results showed increased cerebral blood flow in both groups with similar pain levels, but patients had more regions of activation compared to controls, suggesting that FMS involves augmented pain processing in cortical or subcortical areas.

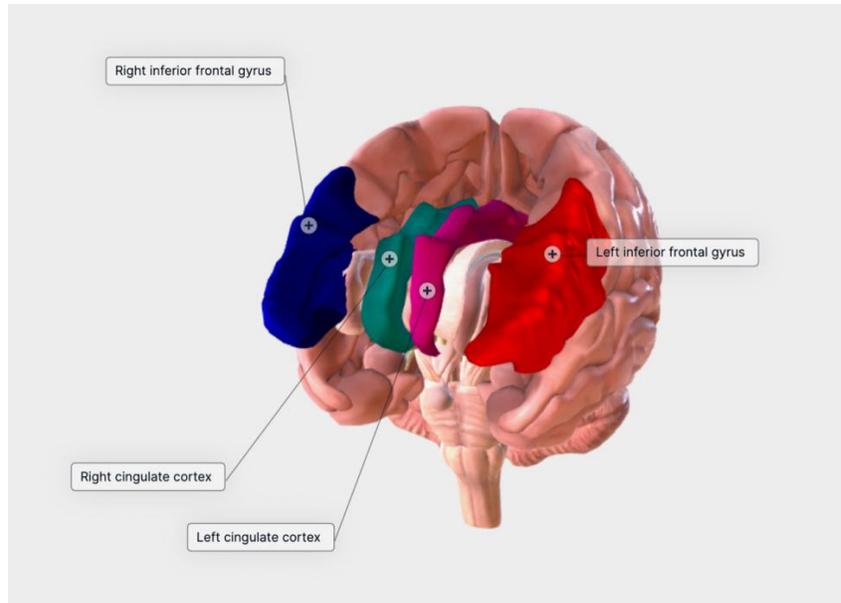

Figure 3  3D Model Images representing bilateral inferior frontal gyri and cingulate cortices, brain structures responsible for pain perception and modulation (created using BioDigital.com)

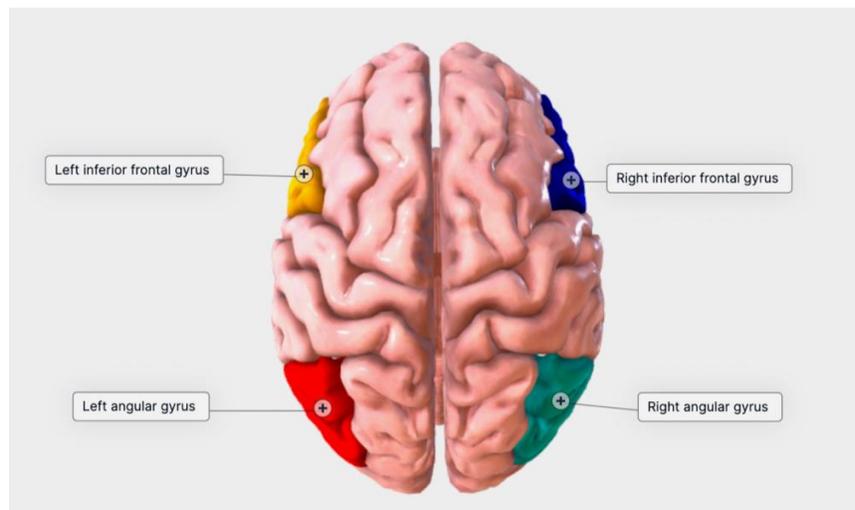

Figure 4  3D Model Images representing bilateral inferior frontal gyri and angular gyri, brain structures responsible for pain perception and modulation (created using BioDigital.com)

   Second, fMRI studies have demonstrated that patients with FMS exhibit heightened pain responses to external stimuli. This increased sensitivity to pain, known as hyperalgesia, is thought to be related to abnormalities in the pain-inhibitory pathways within the central nervous system. Using BOLD fMRI, one study investigated the correspondence between cerebral activity and self-reported pain in patients with FMS. [60] The study involved applying various pressure levels to the thumbnail, ranging from non-painful to painful, and compared cerebral activity between patients with FMS and healthy controls. The findings revealed that brain activity was comparable between the two groups but occurred at lower pressure-pain thresholds in patients with FMS. Another study that used the same approach discovered reduced activity in the rostral anterior cingulate cortex among patients with FMS compared to healthy controls. [61]

Pharmacologic studies of pain response involving fMRI scans before and after treatment with milnacipran, an antidepressant and nerve pain medication, showed notable increases in posterior cingulate cortex activity. These increases were observed in milnacipran responders compared to placebo responders. [59] Interestingly, this increased activity strongly correlated with a reduction in self-reported pain. These findings suggest that milnacipran, a norepinephrine and serotonin reuptake inhibitor, may alleviate fibromyalgia pain by enhancing activity in the posterior cingulate cortex.

Third, fMRI investigations have shed light on the role of central and peripheral contributions in FMS pain. In a study by Staud et al., brain activation patterns in patients with FMS and healthy controls were investigated when exposed to temperature stimuli. [62] Their results revealed numerous non-cortical brain regions, including the thalamus, insula, and cingulate cortex, showing increased activation in response to stimuli. Interestingly, patients with FMS required lower stimulus temperatures to produce equivalent brain activity levels compared to controls. This suggests that the enhanced neural mechanisms in fibromyalgia are not solely due to cortical enhancement. The role of peripheral sensitization in the increased pain sensitivity of patients with FMS was explored by Staud et al. in 2009. Using local anesthetic injections, their findings suggest that peripheral input from muscles may play a larger role in maintaining central sensitization in patients with FMS than was initially expected. [63] Therefore, FMS may involve both peripheral and central contributions, which can vary in their significance among patients.

FMRI provides valuable insights into the neural mechanisms underlying pain processing in FMS. By elucidating the brain's response to pain stimuli in real time, researchers can better understand the pathophysiology of fibromyalgia and potentially identify biomarkers for this condition. This knowledge can aid in the development of targeted treatments, specific diagnostic criteria, and improvement in quality of life for patients with FMS.

**V. The Role of Elastography in Studying Fibromyalgia**

Many non-invasive elasticity imaging techniques exist to measure the mechanical properties of various tissue types quantitatively. [64] These techniques, which include Shear Wave Elastography (SWE) and Magnetic Resonance Elastography (MRE), gather information on tissue elasticity and can be applied to organs located deeper within the body, providing new possibilities for screening and diagnosis. In recent years, emerging elasticity imaging techniques have been developed and studied. [65] Early elasticity imaging techniques from the 1970s and 1980s utilized static loading and external vibration to induce stress in tissues, after which tissue displacement and stiffness were measured using modified color Doppler. In the late 1990s, an alternative, quasi-static approach was developed to remotely measure tissue elasticity through manual compression, now referred to as strain elastography. [66] In this method, the ultrasound probe applies mild pressure to the tissue, causing deformation or compression. The system measures the extent to which the tissue deforms in response to this compression and generates a color-coded map (elastogram) that indicates the relative stiffness of the tissue. See Figure 4. [67] Strain elastography is possible through the varying levels of stiffness throughout the body. For example, healthy tissues tend to be softer and more elastic, whereas diseased or abnormal tissues often become stiffer.

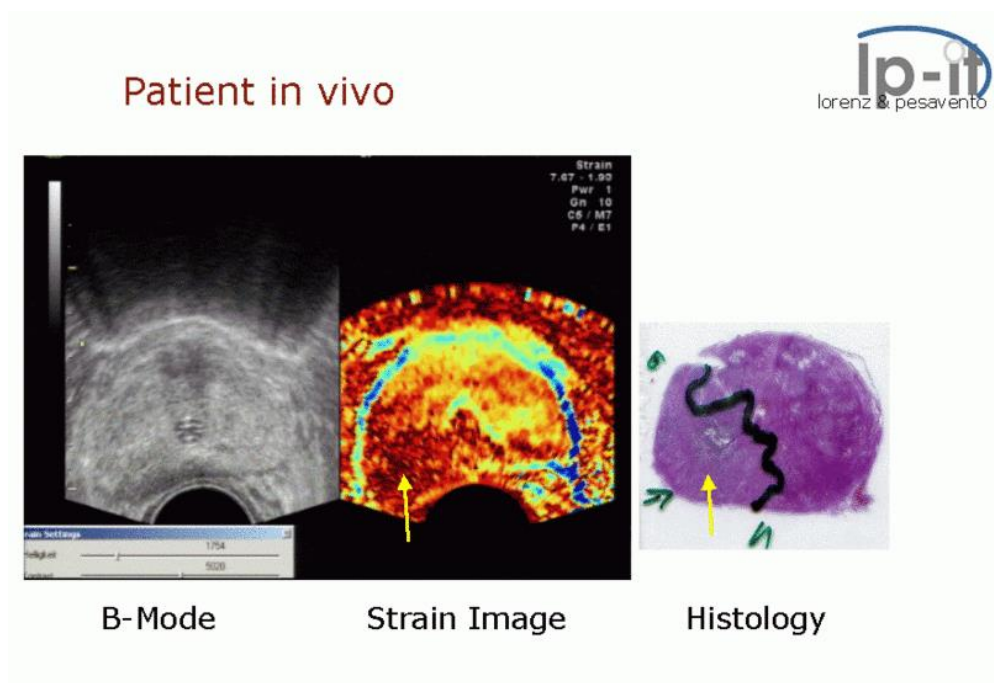

Figure 5  An image showing the detection of prostate cancer by elastography (middle) and histogram (right). A sonographic image is displayed on the left for comparison. The dark, red area at the bottom left of the strain elastography image indicates stiffer tissue. (Andreaslorenzcommon, 2014)

This was followed by the development of dynamic shear-wave elastography, which uses specialized ultrasound transducers to generate shear waves that propagate through tissue. The speed of these shear waves is directly related to tissue stiffness, and by analyzing the propagation speed, the system calculates tissue elasticity and displays it on a quantitative elastogram. [66] Since 2005, more and more manufacturers have incorporated the shear wave elastography option into their standard ultrasound systems, and it is now commonly used in clinical practice to evaluate breast, liver, prostate, and musculoskeletal tissues. [68] However, the measurement of muscle firmness with shear wave ultrasound imaging has never been performed in patients with FMS.

Like SWE ultrasound, MRE makes use of externally applied propagating shear waves, converting them into maps of shear modulus (or stiffness). [69] In place of an ultrasound transducer, an external mechanical driver generates shear waves in a frequency range of 10 – 1000 Hz. These waves are typically imaged with a motion-sensitized MRI pulse sequence, and the resultant wave images are analyzed to obtain quantitative measures of tissue stiffness.

The most widespread application of MRE is the liver, for the identification of hepatic disease. Additional applications of this technique include breast cancer, the brain, and skeletal muscle. [70] Though FMS has not been studied with MRE previously, two MRE studies have examined myofascial taut bands. [71,72]

VI. Other Theoretical Mechanisms of Pain in Fibromyalgia

Muscle nociceptors, also known as muscle pain receptors, are sensory nerve endings found within muscle tissue that respond to noxious or potentially damaging stimuli, leading to pain perception. [73] These specialized nerve endings are sensitive to various types of stimuli, including mechanical pressure, chemical irritants, and extreme temperatures. When muscles are subjected to injury, inflammation, or excessive strain, these receptors are activated and send signals to the brain where the pain sensation is perceived. While these receptors have a high threshold for mechanical stimulation, they are rapidly activated by pain-inducing substances and can convey information about the intensity and duration of noxious stimuli. [74]

Group III and Group IV muscle afferents are two types of sensory nerve fibers found in the peripheral nervous system that transmit information related to function, pain, and reflexes from the muscle tissue to the central nervous system. Group III muscle afferents are myelinated nerve fibers that belong to Aδ fibers. These fast-conducting fibers provide feedback to the central nervous system in terms of muscle length, tension, and joint position, thereby contributing to motor control and coordination. Their main function is to transmit mechanical information and regulate muscle contraction and stretching. Group IV muscle afferents, alternatively, are unmyelinated nerve fibers. Compared with Group III fibers, they conduct nerve signals more slowly. Their functions include conveying nociceptive information and mediating axon reflex. When noxious stimuli activate Group IV muscle afferents, also known as "C fibers," they can trigger a local vasodilation response known as the axon reflex. This response involves the release of vasodilatory neurotransmitters, leading to increased blood flow in the area, and contributes to the inflammatory response and pain sensation. [75]

Consistent with the findings of animal studies, it has been observed that in humans, muscle pain can be induced by applying pressure or chemical stimulation while selectively blocking large-diameter (Groups I and II) afferent fibers. [74] This suggests that muscle pain is mediated by both Group III and Group IV nerve fibers.

Musculoskeletal pain caused by ischemia occurs in a variety of clinical conditions, such as peripheral vascular disease (PVD), sickle cell disease (SCD), complex regional pain syndrome (CRPS), and FMS. [76] Patients with ischemic muscle pain have unique features compared to those with other causes of pain, as they often do not respond well to conventional pain treatments. [77] They also tend to experience heightened cardiovascular responses during muscle contraction, leading to exercise intolerance and exacerbation of adverse cardiovascular conditions. Both Group III and Group IV muscle afferents play a crucial role in the development of pain caused by ischemia and in the sensory component of the exercise pressor reflex (EPR).

EPR is an important autonomic reflex that helps regulate cardiovascular responses during exercise and maintains blood pressure and blood flow to the working muscles. [78] Group III and Group IV afferents are stimulated by muscle contraction, and metabolites such as lactate, hydrogen ions (H+), and potassium ions (K+) accumulate in active muscle tissue. When these muscle afferents are activated, they send signals to the central nervous system and elicit reflexive responses, including increased heart rate and blood pressure.

The development of myalgia and altered cardiovascular function in ischemic conditions involves not only the muscle afferents mentioned above, but also purinergic receptors (P2X and P2Y), transient receptor potential (TRP) channels, and acid-sensing ion channels (ASICs) in peripheral sensory neurons. [76] Specific changes in the primary afferent function through these receptors are associated with heightened pain behaviors and altered EPR responses. Growth factors, cytokines, and microvascular changes in muscles may also contribute to the overexpression of these receptor molecules in the dorsal root ganglia (DRG), further influencing pain and sympathetic reflexes.

The specific causes of deep-tissue pain in FMS are not well understood. However, studies have found evidence of impaired blood flow within the painful areas of the body in patients with FMS. [79,80] Enhanced ultrasound imaging during muscle contractions has shown lower muscle vascularity and reduced flow response in patients with FMS. [80] Additionally, measurements of microcirculation using laser Doppler flowmetry have revealed decreased blood flow above sensitive points in patients with FMS compared to healthy individuals. [79]

**VII.    Future Directions and Research Gaps**

The exact mechanisms linking sustained muscle pressure, chronic pain, and brain changes in fibromyalgia are still being investigated. These theories provide plausible explanations for the observed phenomena but require further research to establish definitive causal relationships.

As of the writing of this review paper, only one study exists comparing muscle pressure in patients with FMS to control subjects. [20] While the findings were significant, using only a single site of muscle pressure (the trapezius) limits the external validity of their results, but provides an opening for future studies. Longitudinal cohort studies following FMS patients, identifying diagnostic biomarkers, comparing the effectiveness of different treatment modalities, evaluating differences in patient outcomes, and investigation of health disparities in fibromyalgia are all potential avenues for future FMS research.

**VIII.    Conclusion**

Fibromyalgia Syndrome is a complex and multifactorial disorder characterized by widespread chronic pain, fatigue, and tenderness at specific anatomical sites. The exact cause of fibromyalgia is not fully understood, and researchers have proposed several theoretical models to explain its etiology, including the Central Sensitization Theory, the Cytokine Inflammation Theory, Muscle Hypoxia, the Muscle Tender Point Theory, and the Small Fiber Neuropathy Theory. Previous studies have linked several important biomarkers to fibromyalgia symptoms; however, whether there is a cause-and-effect relationship remains to be determined. Emphasis was placed on the findings of elevated muscle pressure in fibromyalgia patients, although more research is needed to better understand the underlying mechanisms of this connection. It is important to emphasize that these models and theories are not mutually exclusive, and fibromyalgia is likely a complex interplay of various factors. This review summarizes these different models and expands upon the potential for future research in this field, especially with the advances in fMRI and elastography modalities.

## IX. Funding

All authors declare that they do not have any conflict of interest.